\documentclass[reprint,amsmath,amssymb,
 superscriptaddress,prb,floatfix]{revtex4-1}

\usepackage{graphicx}
\usepackage{chemformula}
\usepackage{float}

\begin{document}
\preprint{APS/123-QED}

\title{Graphene Effusion-based Gas Sensor}

\author{I.E. Ros\l o\'{n}}
\email{E-mail: I.E.Roslon@tudelft.nl}
\affiliation{Kavli Institute of Nanoscience, Delft University of Technology, Lorentzweg 1, 2628 CJ, Delft, The Netherlands}
\affiliation{Department of Precision and Microsystem Engineering, Faculty 3mE, Delft University of Technology, Mekelweg 2, 2628 CD, Delft, The Netherlands}
\author{R.J. Dolleman}
\thanks{Present address: Second Institute of Physics, RWTH Aachen University, Otto-Blumenthal-Stra\ss e, 52074, Aachen, Germany}
\author{H. Licona}
\author{M. Lee}
\author{M. \u{S}i\u{s}kins}
\affiliation{Kavli Institute of Nanoscience, Delft University of Technology, Lorentzweg 1, 2628 CJ, Delft, The Netherlands}
\author{H. Lebius}
\affiliation{CIMAP/GANIL, CEA-CNRS-ENSICAEN-UCN, blvd Henri Becquerel, F-14070 Caen, France} 
\author{\\L. Madau\ss}
\author{M. Schleberger}
\affiliation{Faculty of Physics and CENIDE, Universit\"at Duisburg-Essen, 47057 Duisburg, Germany}
\author{F.Alijani}
\affiliation{Department of Precision and Microsystem Engineering, Faculty 3mE, Delft University of Technology, Mekelweg 2, 2628 CD, Delft, The Netherlands}
\author{H.S.J. van der Zant}
\affiliation{Kavli Institute of Nanoscience, Delft University of Technology, Lorentzweg 1, 2628 CJ, Delft, The Netherlands}
\author{P.G. Steeneken}
\email{E-mail: P.G.Steeneken@tudelft.nl}
\affiliation{Kavli Institute of Nanoscience, Delft University of Technology, Lorentzweg 1, 2628 CJ, Delft, The Netherlands}
\affiliation{Department of Precision and Microsystem Engineering, Faculty 3mE, Delft University of Technology, Mekelweg 2, 2628 CD, Delft, The Netherlands}

\begin{abstract}
Porous, atomically thin graphene membranes have interesting properties for filtration and sieving applications because they can accommodate small pore sizes, while maintaining high permeability. These membranes are therefore receiving much attention for novel gas and water purification applications. Here we show that the atomic thickness and high resonance frequency of porous graphene membranes enables an effusion based gas sensing method that distinguishes gases based on their molecular mass. Graphene membranes are used to pump gases through nanopores using optothermal forces. By monitoring the time delay between the actuation force and the membrane mechanical motion, the permeation time-constants of various gases are shown to be significantly different. The measured linear relation between the effusion time constant and the square root of the molecular mass provides a method for sensing gases based on their molecular mass. The presented microscopic effusion based gas sensor can provide a small, low-power alternative for large, high-power, mass-spectrometry and optical spectrometry based gas sensing methods.
\end{abstract}

\maketitle

\section{Introduction}

Although graphene in its pristine form is impermeable, its atomic thickness causes it to be very permeable when perforated \cite{leenaerts2008graphene, bunch2008impermeable}. This is an advantageous property that has recently been exploited for filtration and separation purposes \cite{joshi2014precise, o2014selective, celebi2014ultimate, kim2013selective}. For sub-nm pore sizes, it has been shown to result in molecular sieving \cite{sint2008selective, cohen2012water} and osmotic pressure \cite{dolleman2016graphene}. Besides filtration and separation, selective permeability might also provide a route toward sensing applications. In contrast to chemical \cite{yuan2013graphene} and work-function based \cite{chatterjee2015graphene} gas sensing principles, permeation does not rely on chemical or adhesive bonds of the gas molecules, which can be irreversible or require thermal or optical methods to activate the desorption of the bound gas molecules \cite{wang2016review}. A permeation based gas sensor can feature improved response-time, robustness and lifetime, and can enable sensing of inert gases. 

\begin{figure}[ht!]
  \includegraphics{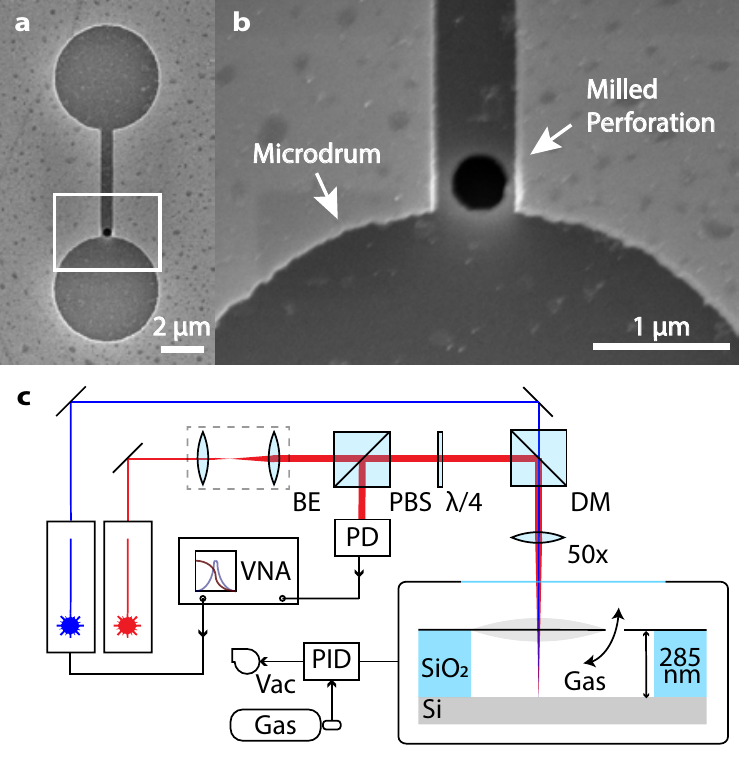}
  \caption{a) SEM image of dumbbell shaped cavities covered by bilayer graphene. b) A nanopore with a diameter of 400 nm is milled in the graphene by FIB in the channel that connects the two microdrums. c) Interferometry setup used to actuate and detect the motion of the graphene microdrums. The red laser passes subsequently through the beam expander (BE), the polarized beam splitter (PBS) and the quarter-wave plate ($\lambda/4$), after which it is combined with the blue laser using a dichroic mirror (DM) and focused on the drum using a 50x objective. The readout is performed by a high frequency photodiode (PD) that is connected to the Vector Network Analyzer (VNA). The VNA modulates the power of the blue laser that actuates the membrane. Gas pressure inside the vacuum chamber is controlled by a PID controller.} 
  \label{fgr:MeasSetup}
\end{figure}

When gas molecules flow through pores that are smaller than the mean free path length, but larger than their kinetic diameter, their permeation is in the effusive regime. According to Graham's law \cite{maxwell1860v}, the effusion time constant is proportional to the square root of the gas molecular mass. Here, we demonstrate that these effusion effects can be utilized for permeation based gas sensing. By using graphene membranes to pump gases \cite{davidovikj2018graphene} through focus ion beam (FIB) milled nanopores \cite{schleberger20182d}, we realize a fast, low-power and miniaturizable gas sensor. The permeation rate is determined from the frequency ($\omega$) dependent response function $z_\omega/F_\omega$ which is used to determine the gas-specific time-delay $\tau_{gas}$ between the optothermal actuation force $F_\omega$ and the membrane displacement $z_\omega$. We show that the permeation time-constants can be engineered by altering the number of pores, their cumulative area and by adding a flow resistance in the form of a gas channel in series with the pore. \\

Figure \ref{fgr:MeasSetup}a and \ref{fgr:MeasSetup}b show a scanning electron microscope (SEM) top-view of the sensor device. Dumbbell-shaped cavities are etched in a silicon substrate with a 285 nm SiO$_2$ layer using reactive ion etching, creating drums with a diameter of 5 $\mu$m that are connected by a channel of 0.6 $\mu$m wide and 5 $\mu$m long \cite{davidovikj2018graphene}. A stack of two chemical vapour deposited (CVD) monolayers of graphene is transferred over the cavity with a dry transfer method by Applied Nanolayers B.V. and subsequently annealed in an argon furnace. Nanoscale circular pores with diameters varying from 10 nm to 400 nm are milled through the suspended CVD graphene using FIB \cite{celebi2014ultimate}. Pores are created in the channel instead of the drum, as directly milling on the drum reduced signal quality.

The frequency response curves of the membranes are measured using an interferometry setup shown in Figure  \ref{fgr:MeasSetup}c. The setup consists of two lasers that are focused with a 1.5 $\mu$m spot size on the sample in the vacuum chamber. The red laser ($\lambda_{\rm red}=632.8$ nm) is used for detection of the amplitude and phase of the mechanical motion, where the position-dependent optical absorption of the graphene results in an intensity modulation of the reflected red laser light, that is detected by a photodiode \cite{castellanos2013single}. A power-modulated blue laser ($\lambda_{\rm blue}=488$ nm), which is driven by a vector network analyzer (VNA) at frequencies from 9 kHz to 100 MHz, optothermally actuates the membrane motion \cite{dolleman2018opto}. The incident red and blue laser powers are 2 mW and 0.3 mW, respectively. A calibration measurement, in which the blue laser is directly illuminating the photodiode, is used to eliminate systematic parasitic delays in the system \cite{dolleman2017optomechanics}. 

\section{Operation principle}

We now discuss how the frequency dependent mechanical response of the graphene drum to the modulated laser actuation can be used to characterize the gas permeation rate through the porous membranes. In vacuum, the graphene membrane is solely actuated by thermal expansion, as a consequence of the temperature variations induced by the modulated blue laser. This effect has been extensively studied by Dolleman \emph{et al.} \cite{dolleman2017optomechanics} to characterize the heat transport from membrane to substrate. The temperature at the center of the membrane $T(t) = T_{\rm ext} + \Delta T$, where $T_{\rm ext}$ is the ambient temperature, can be approximately described by a first order heat equation, where the optothermal laser power $\mathcal{P}_{\rm AC} e^{i \omega t}$ is absorbed by the graphene membrane and thermal transport towards the substrate is approximated by a single thermal time constant $\tau_{\rm th}=R_{\rm th} C_{\rm th}$ corresponding to the product of the membrane's thermal resistance and thermal capacitance: 
\begin{equation}
\label{eq:heateq}
\frac{{\rm d}\Delta T}{{\rm d} t}  = -\frac{\Delta T}{\tau_{\rm th} } + \frac{\mathcal{P}_{\rm AC}}{C_{\rm th}} e^{i \omega t}.
\end{equation}
In the presence of gas, the pressure difference $\Delta P=P-P_{\rm ext}$ between the cavity pressure $P$ and the ambient pressure $P_{\rm ext}$ can also be described by a differential equation. There are three contributions to the time derivative of the pressure ${\rm d}\Delta P/{\rm d}t$: gas permeation, motion of the membrane and laser heating of the gas in the cavity: 
\begin{equation}
\label{eq:pdiff}
\frac{{\rm d}\Delta P}{{\rm d}t} = -\frac{\Delta P}{\tau_{\rm gas} } + \gamma \frac{{\rm d}z}{{\rm d}t} +  \frac{\mathcal{P}_{\rm AC}}{C_{\rm gas}} e^{i \omega t}.
\end{equation}
Gas permeation out of the membrane with a time constant $\tau_{\rm gas}$ gives a contribution $-\Delta P/\tau_{\rm gas}$. Compression of the gas by the downward deflection $z$ of the membrane results in a term $\gamma {\rm d}z/{\rm d}t$, where $\gamma$ is a constant of proportionality. Heating of the gas due to power absorption of the modulated laser can be described by a term $ \frac{\mathcal{P}_{\rm AC}}{C_{\rm gas}} e^{i \omega t}$, where $C_{\rm gas}$ is a constant relating thermal power to gas expansion. 

A third differential equation is used to describe the mechanics of the membrane, which at low amplitudes experiences a force contribution proportional \cite{dolleman2017optomechanics,davidovikj2017nonlinear} to the pressure difference $F_P=\beta \Delta P$ and an effective thermal expansion force $F_T=\alpha \Delta T$: 
\begin{equation}
\label{eq:dyn}
m_{\rm eff}\frac{{\rm d}^2z}{{\rm d}t^2} + c \frac{{\rm d}z}{{\rm d}t} + kz = \alpha \Delta T + \beta \Delta P.
\end{equation}
Here, we describe the fundamental mode of motion at the center of the membrane by a single degree of freedom forced harmonic oscillator with effective mass $m_{\rm eff}$. The resulting system of three differential equations (\ref{eq:heateq}-\ref{eq:dyn}) is solved analytically for frequencies significantly below the resonance frequency $\omega_{\rm res} = \sqrt{k/m_{\rm eff}}$, where terms proportional to ${\rm d}^2z / {\rm d}t^2$ and ${\rm d}z / {\rm d}t$ can be neglected, to obtain the complex frequency response $z_{\omega} / \mathcal{P}_{\rm AC}$ of the membrane. A full derivation, solution and numerical simulation of the three differential equations can be found in the Supplementary Information 1 and 2. The real and imaginary parts of the solution relate to the components of the displacement $z_\omega$ that are in-phase and out-of-phase with respect to the laser power modulation. The imaginary part of this expression is found to be:
\begin{equation}
Im(z_{\omega}) = a\frac{\omega\tau_{\rm th}}{1+ \omega^2 \tau_{\rm th}^2}   +   b\frac{\omega\tau_{\rm gas} }{1+ \omega^2 \tau_{\rm gas}^2}
\label{eq:ImagExpression}
\end{equation}
 This equation is used to fit to the experimental data with $a$, $b$, $\tau_{\rm th}$ and $\tau_{\rm gas}$ as fit parameters. At frequencies close to the reciprocal permeation time $\omega_{\rm gas} = 1/\tau_{\rm gas}$ the imaginary part of the displacement displays a minimum, similar to the effect observed near $\omega_{\rm th} = 1/\tau_{\rm th}$ for the thermal actuation \cite{dolleman2017optomechanics}. In the following, these extrema in the imaginary part of the frequency response will be used for characterizing permeation and thermal time-constants.

\section{Results}
\begin{figure}
  \includegraphics[width=0.8\linewidth]{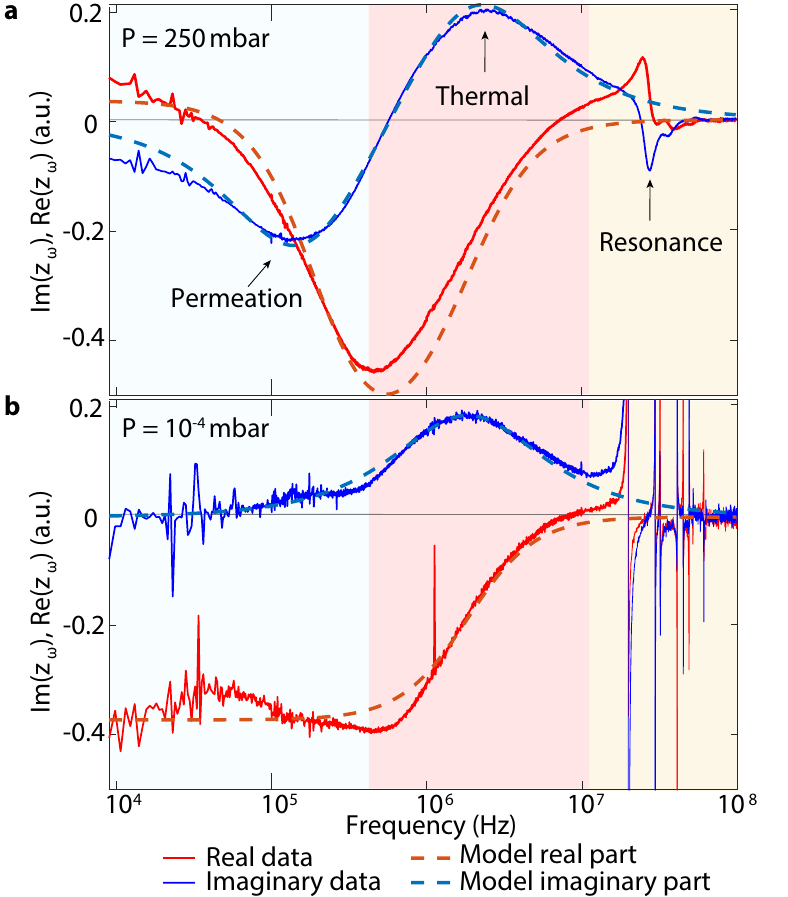}
  \caption{a) Frequency response of the device shown in Figure \ref{fgr:MeasSetup} in nitrogen gas at $P = 250$ mbar showing the real (in phase) and imaginary (90 deg phase shift) parts of the signal, $\operatorname{Im}(z_{\omega})$, $\operatorname{Re}(z_{\omega})$ b) Measurement on the same device at $P < 10^{-4}$ mbar shows that the permeation peak disappears in vacuum and the thermal peak remains almost unchanged.}
  \label{fgr:Response}
\end{figure}

A typical frequency response curve $z_\omega$ of a device at a pressure $P = 250$ mbar in nitrogen gas is shown in Figure \ref{fgr:Response}a. The mechanical resonance occurs in the MHz domain, here at $f = 25.9$ MHz with $Q = 4.2$. Below the mechanical resonance, the imaginary response $\operatorname{Im}(z_{\omega})$ shows two characteristic peaks at 160 kHz and 2 MHz, which are assigned to the extrema of equation (\ref{eq:ImagExpression}) corresponding to fit parameters $\tau_{\rm th} = 1/(2 \pi \cdot 2 \: {\rm MHz}) = 81$ ns and $\tau_{\rm gas} = 1/(2 \pi \cdot 160 \:{\rm kHz}) = 991$ ns.

To prove that one of the peaks is related to gas permeation, we repeat the measurement in vacuum. The measurements in high vacuum show only one maximum in the imaginary response, corresponding to a thermal time $\tau_{\rm th} = 87$ ns, as shown in Figure \ref{fgr:Response}b. Also, reference samples without perforations show only one peak with a similar time-constant $\tau_{\rm th}$. Therefore, it is concluded that the peak in $\operatorname{Im}(z_{\omega})$ at 160 kHz in Figure \ref{fgr:Response}a is due to gas permeation. 

\begin{figure}
  \includegraphics[width=1\linewidth]{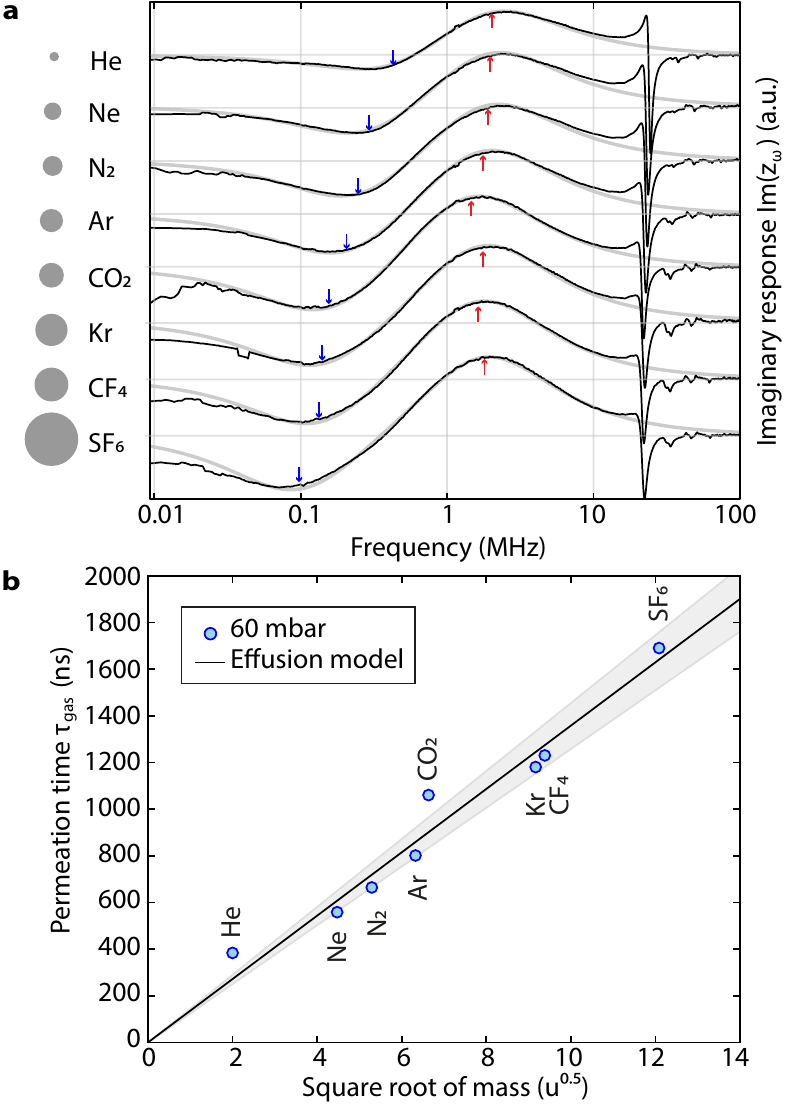}
  \caption{a) Measurements performed in various gases at $P = 60$ mbar on a device with a 400 nm pore. The areas of the circles represent the relative mass of the gas particles. b) The permeation time $\tau_{\rm gas}$ increases linearly with the square root of the particle mass as predicted by classical effusion.}
  \label{fgr:Results}
\end{figure}

The permeation time constants $\tau_{\rm gas}$ are extracted for a range of gases varying in molecular mass $M$ from 4 u (He) to 130 u (\ch{SF6}). Figure \ref{fgr:Results} shows that the permeation time constant closely follows Graham's effusion law with $\tau_{\rm gas} \propto \sqrt[]{M}$. The slope of the linear effusion model is fitted to the data, and the grey area shows the 95\% confidence interval. This agreement demonstrates that the porous graphene membranes can be used to distinguish gases based on their molecular mass. A significant deviation between measurement and theory is only observed for He, which could be due to fitting inaccuracies related to the proximity of the thermal time-constant and mechanical resonance frequency peaks to the gas permeation related peak. 

The gas permeation time $\tau_{\rm gas}$ can be tuned by varying the cumulative pore area, either by changing the number of pores or their size. This tuning can be useful, since too short time constants will lead to overlap between the $\tau_{\rm gas}$ and $\tau_{\rm th}$ peaks or even with the resonance peaks, whereas long permeation rates could be problematic in view of acquisition times. 

\begin{figure}
  \includegraphics[width=1\linewidth]{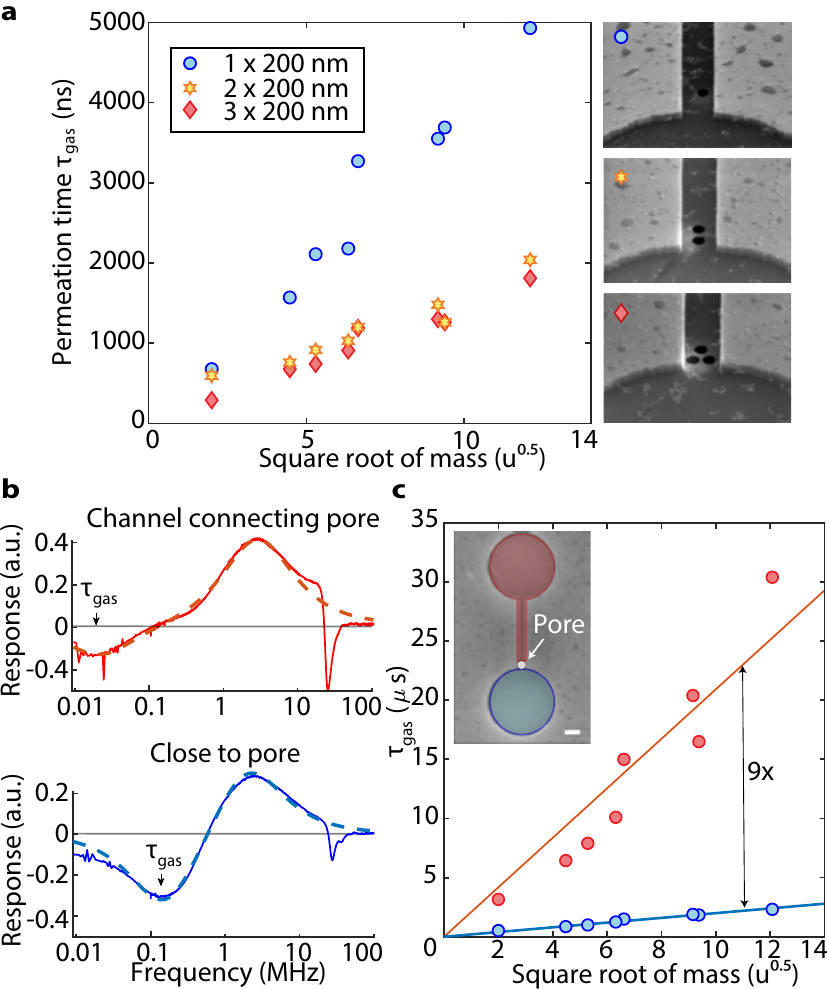}
  \caption{a) Comparison of measurements between three different devices with increasing number of 200 nm pores. The permeation time $\tau_{\rm gas}$ reduces with the cumulative pore area $A$, but saturates at 3 pores. b) Measurements of $\operatorname{Im}(z_\omega)$ with the laser aimed at the drum next to the pore and at the the drum connected by a channel to the pore, respectively the blue and the red drum in the SEM inset in c, showing a large tuning of $\tau_{\rm gas}$. c) The gas permeation time of the drum close to the perforation is 9 times shorter than of the drum far away from the perforation. Inset: SEM image (false color) of the two graphene drums connected by a channel with a 400 nm circular pore, scalebar $= 1 \:\mu$m. All measurements in this figure are performed at $P = 60$ mbar.}
  \label{fgr:Tuning}
\end{figure}

Figure \ref{fgr:Tuning}a demonstrates $\tau_{\rm gas}$ tuning in devices with increasing number of 200 nm pores. The permeation time $\tau_{\rm gas}$ decreases with the cumulative pore area $A$ when increasing the number of pores. The average reduction of $\tau_{\rm gas}$ by a factor $2.26 \pm 0.5$ when doubling the number of pores from 1 to 2 is additional evidence that this time-constant is related to the permeation rate. The change in the permeation time by a factor higher than two when doubling the number of pores might be caused by the fact that the two pores are located closer to the drum than the single pore, leading to a higher transmission probability. When increasing the number of pores to 3, the time-constant does not drop accordingly, indicating that other effects than pore effusion limit the permeation rate, such as the time taken by the gas to reach the pores at the edge of the drum. 

We investigate this gas sensing approach further by tuning $\tau_{\rm gas}$ by placing the holes further away from the graphene drum, at the other end of the channel that connects both drums. The SEM inset of Figure \ref{fgr:Tuning}c shows a pore inside the channel, that is close to the blue drum, but far from the red drum. The rectangular, graphene-covered channel, with dimensions of $5\times 0.6\times 0.285$ $\mu{\rm m}^3$, is in series with the pore for the red drum. It is found from Figure \ref{fgr:Tuning}c that the permeation time is 9 times longer for the red drum that is in series with the channel. The difference in permeation time is a measure of the transmission probability $\psi_r$ through the rectangular channel. In the ballistic regime, the conductance and time-constant are given as the product of the time-constant of the aperture (the pore) and the transmission probability of the channel $\psi_r$ meaning that $\tau_{\rm gas, close} = \psi_r \times \tau_{\rm gas, far}$. The transmission probability through a rectangular channel can be calculated using the Smoluchowski formula \cite{smoluchowski1910kinetischen} for which an useful approximation \cite{clausing1971flow,livesey1998flow} is given by:
\begin{equation}
    \psi_r = \frac{16}{3\pi^{3/2}} \frac{a}{l} \mathrm{ln}\left( 4\frac{b}{a} + \frac{3}{4}\frac{a}{b} \right),
\end{equation}
where $a = 285$ nm and $b = 600$ nm are the cross-sectional dimensions and $l = 5 \: \mu$m is the length in the direction of gas flow. The formula predicts a 12\% transmission probability for our geometry, in close agreement with the experimental value of 11\% that is found from the ratio between the slopes of the blue and red solid lines in Figure \ref{fgr:Results}.

\begin{figure*}
  \includegraphics{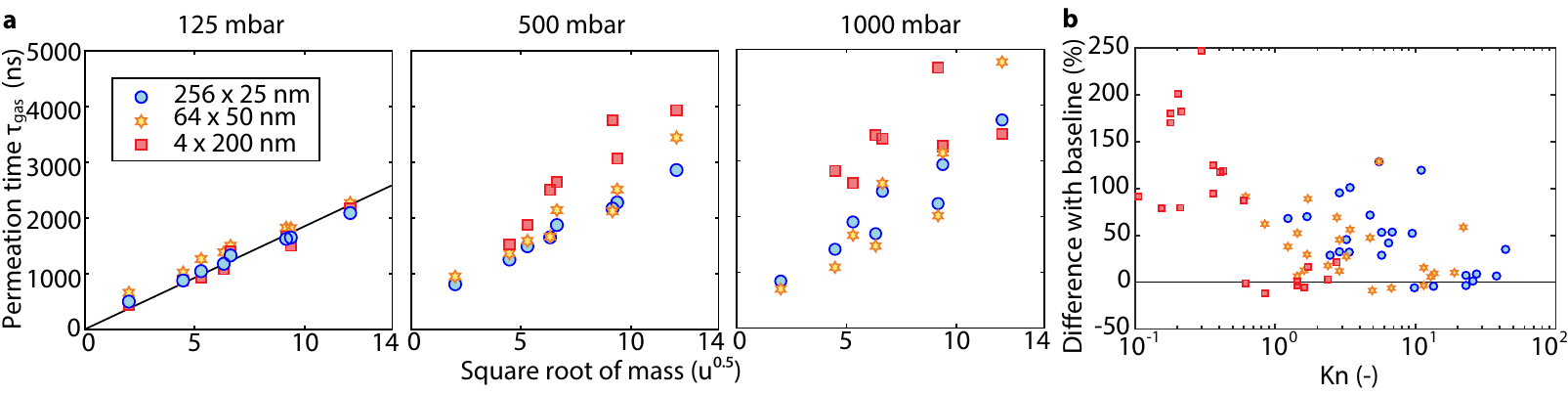}
  \caption{Permeation times in the transitional region between Knudsen and slip flow at a Kn range of 0.1-10. As Kn drops below 1, the permeation time increases and Graham's law does no longer describe the values of $\tau_{\rm gas}$ correctly.}
  \label{fgr:Poresize}
\end{figure*}

The size of individual pores determines whether viscous Sampson or molecular Knudsen flow is taking place \cite{boutilier2017knudsen}. Figure \ref{fgr:Poresize} compares time-constants $\tau_{\rm gas}$ in devices with equal cumulative area $A = 4 \pi \cdot 10^4$ nm$^2$ and different pore diameters. At $P=125$ mbar all devices show a linear relation between the square root of mass and the permeation time according to Graham's law. In contrast, at higher pressures where the free path length $\lambda$ becomes smaller than the pore diameter $d$ (Kn = $\lambda/d<1$), in particular for the larger molecular masses and large pore sizes, the linear dependence disappears. In the transitional region between Knudsen and Sampson flow, classical effusion no longer correctly describes the flow and viscosity effects lead to larger values of $\tau_{\rm gas}$ than predicted by Graham's law. This increase is in line with studies on pipe and channel flows, \cite{tison1993experimental} which show a maximum in the permeation time near Kn=1 where the transition from Knudsen to Sampson flow occurs. 

Besides permeation, gas sensing can be achieved by observing changes in the thermal time constant in a fashion similar to Pirani gas sensors. In general the gas conducts heat better at higher pressures, and it does so also for molecules with a smaller molecular mass and higher molecular velocity. However, by analyzing the values of $\tau_{\rm th}$ that are determined from measurements like in figure \ref{fgr:Results}a, it appears that thermal conductivity of the gases is a less precise route toward gas sensing than the permeation based method shown in Figure \ref{fgr:Results}b. Further experimental results for the thermal time constant can be found in the supplementary information.

\section{Conclusion and Discussion}
In conclusion, a gas sensing method is presented based on measuring the permeation time-constant $\tau_{\rm gas}$ of gases through pores in bilayer graphene membranes. Due to the small pore sizes, permeation is governed by effusion, such that permeation rates are inversely proportional to the square root of the molecular mass of the gas. By optothermal driving, the gas in the cavity below the graphene membrane is pressurized and pumped through the porous membrane. At angular driving frequencies close to the inverse of the permeation time constant ($\omega = 1/\tau_{\rm gas}$), a peak in the imaginary part of the frequency response appears which is used to characterize the gas species based on their effusion speed. By changing the number of pores and pore diameter using FIB, the time constants can be adjusted to the desired range, which might be achieved by sensor arrays of different dimensions and pore sizes. For practical sensing applications, however, the precision, signal to noise, speed and readout protocol for $\tau_{\rm gas}$ will need to be further improved. Besides presenting a permeation based gas sensing concept, this work also shows that the extreme flexibility and permeability of suspended porous membranes of 2D materials can be used as an interesting platform for studying thermodynamics of gases at the nanoscale.

\begin{acknowledgements}

The authors thank Applied Nanolayers B.V. for the supply and dry transfer of bilayer graphene. This work is part of the research programme Integrated Graphene Pressure Sensors (IGPS) with Project Number 13307 which is financed by The Netherlands Organisation for Scientific Research (NWO). The research leading to these results also received funding  from the European Union’s Horizon 2020 research and innovation program under Grant Agreement No. 785219 Graphene Flagship and within the FLAG-ERA project NU-TEGRAM. M.S. and L.M. acknowledge funding by the Deutsche Forschungsgemeinschaft (Project C5 within the SFB 1242 “Non-Equilibrium Dynamics of Condensed Matter in the Time Domain” (project No. 278162697) and SCHL 384/16-1 (project No.  279028710). 

\end{acknowledgements}

\clearpage
\onecolumngrid
\appendix

\section*{Supplementary Material: Graphene Effusion-based Gas Sensor}
\section{Model derivation}

In this section we derive the model for the complex amplitude of the membrane. The temperature at the center of the membrane $T(t)$ can be approximately described by a first order heat equation: 
\begin{equation}
\label{eq:heateq1}
\frac{{\rm d}\Delta T}{{\rm d} t}  = -\frac{\Delta T}{\tau_{\rm th} } + \frac{\mathcal{P}_{\rm AC}}{C_{\rm th}} e^{i \omega t},
\end{equation} 

where the optothermal laser power $\mathcal{P}_{\rm AC} e^{i \omega t}$ is absorbed by the graphene membrane and thermal transport towards the substrate is determined by a single thermal time constant $\tau_{\rm th}=R_{\rm th} C_{\rm th}$ corresponding to the product of the membrane's thermal resistance and thermal capacitance.

In the presence of gas, the pressure difference $\Delta P=P-P_{\rm ext}$ between the cavity pressure $P$ and the ambient pressure $P_{\rm ext}$ can also be described by a differential equation. There are three contributions to the time derivative of the pressure ${\rm d}\Delta P/{\rm d}t$: gas permeation, motion of the membrane and laser heating of the gas in the cavity. 
\begin{equation}
\label{eq:pdiff1}
\frac{{\rm d}\Delta P}{{\rm d}t} = -\frac{\Delta P}{\tau_{\rm gas} } + \gamma \frac{{\rm d}z}{{\rm d}t} +  \frac{\mathcal{P}_{\rm AC}}{C_{\rm gas}} e^{i \omega t}
\end{equation}
Gas permeation out of the membrane with a time constant $\tau_{\rm gas}$ gives a contribution $-\Delta P/\tau_{\rm gas}$. Compression of the gas by the downward deflection $z$ of the membrane results in a term $\gamma {\rm d}z/{\rm d}t$, where for small $z$ and cavity depth $g$, it can be shown from Boyle's law that $\gamma = \eta P_{ext}/g$, where $\eta$ is a factor that depends on the deformed shape of the membrane ($\eta=1$ for a piston like membrane motion). Heating of the gas due to power absorption of the modulated laser can be described by a term $ \frac{\mathcal{P}_{\rm AC}}{C_{\rm gas}} e^{i \omega t}$, where $1/C_{\rm gas}={\rm d}P/{\rm d}U$ is the pressure increase per absorbed laser heat energy $U$. For a gas at constant volume $V$, the temperature induced pressure change is given by the ideal gas law as ${\rm d}T/{\rm d}P = \frac{V}{N k_B}$, where $N$ is the number of gas molecules and $k_B$ is Boltzmann's constant. The temperature change for a certain absorbed amount of heat is given by ${\rm d}U/{\rm d}T = c_v m$, where $c_v$ is the specific heat and $m$ the mass of the gas molecules. Thus it is found that the power induced gas pressure increase is characterized by the constant $C_{\rm gas}={\rm d}U/{\rm d}T \times {\rm d}T/{\rm d}P = V m c_v / N K_B $. 

A third differential equation is used to describe the mechanics of the membrane, which at low amplitudes experiences a force contribution proportional \cite{davidovikj2017nonlinear} to the pressure difference $F_P=\beta \Delta P$ and an effective thermal expansion force $\alpha \Delta T$. We approximate the fundamental mode of motion of the center of the membrane by a forced harmonic oscillator with effective mass $m_{eff}$ to obtain: 
\begin{equation}
\label{eq:dyn1}
m_{eff}\frac{{\rm d}^2z}{{\rm d}t^2} + c \frac{{\rm d}z}{{\rm d}t} + kz = \alpha \Delta T + \beta \Delta P.
\end{equation}
 The resulting system of 3 differential equations (\ref{eq:heateq1}-\ref{eq:dyn1}) is solved analytically for frequencies below the resonance frequency, where terms proportional to ${\rm d}^2z/{\rm d}t^2$ and ${\rm d}z/{\rm d}t$ can be neglected, to obtain the complex frequency response of the membrane. For frequencies well below the resonance frequency the induced amplitude can be approximated by:
\begin{equation}
z_{\omega} \approx \alpha \Delta T_{\omega} + \beta \Delta P_{\omega}.
\label{complexamplitude}
\end{equation}
This can be substituted into equation \ref{eq:pdiff1} to arrive at: 
\begin{equation}
\frac{d\Delta P}{dt} + \frac{\Delta P}{(1- \beta \gamma)\tau_{\rm gas}} = \frac{\gamma \alpha}{(1 - \beta \gamma)} \frac{d\Delta T}{dt} + \frac{\mathcal{P}_{\rm AC}}{(1- \beta \gamma)C_P} e^{i \omega t}.
\end{equation}
This expression still depends on the temperature $\Delta T$ of the membrane. A solution to the temperature $\Delta T$ of the membrane following equation 2 in the main text, as found by Dolleman \emph{et al.}, is given by:
\begin{equation}
\Delta T_{\omega} = \frac{R_{th}\mathcal{P}_{AC}}{i \omega \tau_{th} + 1} e^{i \omega t}.
\label{eqTempExpr}
\end{equation}
This solution is used to arrive at:
\begin{equation}
\frac{d\Delta P}{dt} + \frac{\Delta P}{(1- \beta \gamma)\tau_{\rm gas}} = \frac{\gamma \alpha R_{\rm th} \mathcal{P}_{\rm AC}}{(1 - \beta \gamma)} \frac{i \omega e^{i \omega t}}{i \omega \tau_{\rm th} +1} + \frac{\mathcal{P}_{\rm AC}}{(1- \beta \gamma)C_P} e^{i \omega t}.
\end{equation}
Next, we assume $\gamma = 0$, which holds true for small membrane deflections. We now arrive at:
\begin{equation}
    \frac{d\Delta P}{dt} + \frac{\Delta P}{\tau_{\rm gas}} = \frac{\mathcal{P}_{\rm AC}}{C_P} e^{i \omega t}.
\end{equation}
By solving this differential equation a solution for $\Delta P$ is found:
\begin{equation}
\Delta P_{\omega} = \frac{\tau_{\rm gas}}{C_P} \frac{\mathcal{P}_{AC}}{i \omega \tau_{gas} + 1} e^{i \omega t}.
\label{eqPresExpr}
\end{equation}
By inserting expressions \ref{eqTempExpr} and \ref{eqPresExpr} into formula \ref{complexamplitude}, the complex amplitude $z_{\omega}$ can be obtained:
\begin{equation}
z_\omega e^{i \omega t} =  \frac{\alpha R_{\rm th}\mathcal{P}_{\rm AC}}{i \omega \tau_{\rm th} + 1} e^{i \omega t} + \frac{ \tau_{\rm gas}}{C_P} \frac{\beta\mathcal{P}_{\rm AC}}{i \omega \tau_{\rm gas} + 1} e^{i \omega t}.
\label{eqFullExpression}
\end{equation}
The imaginary part of the complex amplitude is calculated:
\begin{equation}
\begin{aligned}
{\rm Im}(z_{\omega}) = \frac{\alpha \tau_{\rm th} R_{\rm th} \mathcal{P}_{\rm AC} }{1+ \omega^2 \tau_{\rm th}^2}   +  \frac{ \tau_{\rm gas}}{C_P} \frac{\beta \tau_{\rm gas} \mathcal{P}_{\rm AC} }{1+ \omega^2 \tau_{\rm gas}^2}.
\end{aligned}
\label{eqImagExpression}
\end{equation}
This is the same equation 4 in the main text that is used for fitting, where $a = \alpha R_{\rm th} \mathcal{P}_{\rm AC}$ and $b =  \frac{ \beta \tau_{\rm gas} \mathcal{P}_{\rm AC} }{C_P}$.

\section{Numerical simulation}

\begin{figure}
  \includegraphics[width=0.6\linewidth]{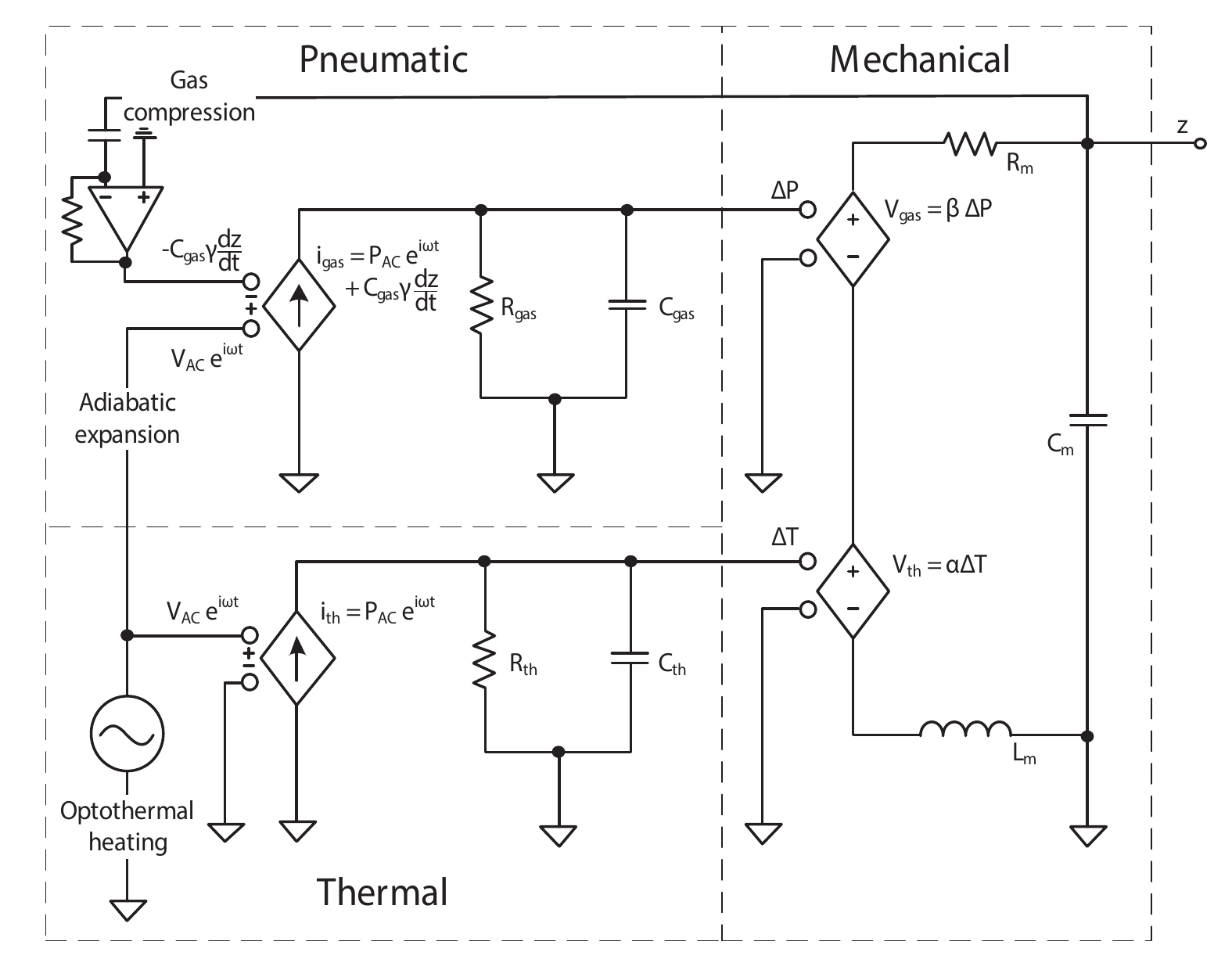}
  \centering
  \caption{Equivalent electric model for the porous membrane.}
  \label{fgr:Electric}
\end{figure}

The system of 3 differential equations (\ref{eq:heateq1}-\ref{eq:dyn1}) is numerically simulated using an analogy to the currents running in an electric circuit. The circuit consists of a thermal, a mechanical and a pneumatic domain, as shown in Figure \ref{fgr:Electric}. The domains are discussed one by one. Simulations have been performed using Simulink.
\paragraph{Mechanical}
The mechanical motion of the membrane is represented by a driven damped harmonic oscillator. The equation of motion for the membrane is represented by an RLC circuit in figure \ref{fgr:Electric} with a resistor $R_m = c$, an inductor $L_m = m$ and a capacitor $C_m = 1/k$, driven by two voltage controlled voltage sources, $V_{th} = \alpha \Delta T$ and $V_{gas} = \beta \Delta P$. The equation of motion is written next to the expression for the electric potential in this circuit:
\begin{equation*}
\begin{aligned}
m\frac{d^2z}{dt^2} + c \frac{dz}{dt} + kz &= \alpha \Delta T + \beta \Delta P \\
L_m\frac{d^2q}{dt^2} + R_m \frac{dq}{dt} + \frac{q}{C_m} &= V_{\rm th} + V_{\rm gas}
\end{aligned}
\end{equation*}
Comparison shows that the charge $q$ on the capacitor in this circuit can represent the deflection $z$ of the membrane. In the schematic the voltage over the capacitor, $V_C = \frac{q}{C_m}$, is taken as an output for readout. 

\begin{figure}
  \includegraphics[width=0.6\linewidth]{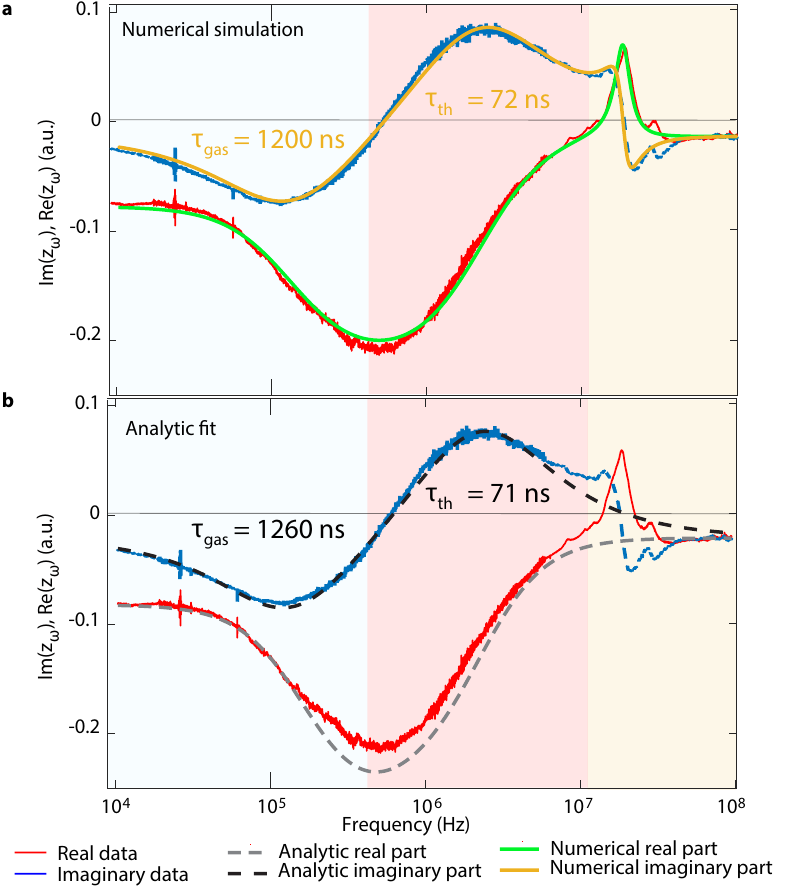}
  \centering
  \caption{a) The governing differential equations (\ref{eq:heateq1} - \ref{eq:dyn1}) are solved numerically using Simulink, yielding a nearly perfect fit to the measured frequency response curves of the porous membrane including the first resonance peak. b) A fit using the analytic formula, yielding nearly the same fitting parameters as with the numerical simulation.}
  \label{fgr:sim}
\end{figure}

\paragraph{Thermal}
The optothermal drive actuating the membrane is represented by an AC voltage source. It controls the voltage controlled current source driving a parallel RC circuit, resembling the thermal flux delivered to the graphene with heat capacity $C_{\rm th}$ and thermal boundary resistance $R_{\rm th}$. The equation for the membrane temperature is written next to the equation for the currents running through this circuit:
\begin{equation*}
\begin{aligned}
\frac{d\Delta T}{dt} + \frac{\Delta T}{\tau_{\rm th} } &= \frac{\mathcal{P}_{\rm AC}}{C_{\rm th}} e^{i \omega t} \\
C_{\rm th}\frac{dV_C}{dt} + \frac{V_C}{R_{\rm th}} &= i_{\rm th}
\end{aligned}
\end{equation*}
Comparison shows that the voltage across the capacitor $V_C$ can represent the temperature of the membrane $T$. Thermal expansion sets the membrane in motion. Therefore, this voltage controls the source driving the circuit in the mechanical domain.

\paragraph{Pneumatic}
The optothermal drive causing adiabatic expansion of the gas is represented by an AC voltage source. Moreover, the movement of the membrane compresses the gas. The voltage over the capacitor in the mechanical domain $kz$ controls a voltage controlled voltage source which is connected to a derivator to change the signal into the effective compression $-C_{\rm gas} \gamma \frac{dz}{dt}$. A voltage controlled current source drives an RC circuit consisting of a capacitor $C_{\rm gas}$ and a resistor $R_{\rm gas}$ in parallel. This circuit resembles the pressure in the cavity with corresponding effective pressure capacity and permeation resistance. The equation for the pressure in the cavity is written next to the equation for the currents running through this circuit:
\begin{equation*}
\begin{aligned}
\frac{d\Delta P}{dt} + \frac{\Delta P}{\tau_{\rm gas} } &= \frac{\mathcal{P}_{\rm AC}}{C_{\rm gas}} e^{i \omega t} + \gamma \frac{dz}{dt} \\
C_{\rm gas}\frac{dV_C}{dt} + \frac{V_C}{R_{\rm gas}} &= i_{\rm gas} + \gamma \frac{dV_z}{dt}
\end{aligned}
\end{equation*}
Comparison shows that the voltage across the capacitor $V_C$ can represent the pressure inside the cavity $P$. The force exerted by the gas sets the membrane in motion. Therefore, this voltage controls the source driving the circuit in the mechanical domain.

The frequency response of a device is numerically simulated. Figure \ref{fgr:sim}a compares shows data from a device in nitrogen gas with 64 $\times$ 50 nm pores with fitting parameters $\tau_{\rm gas} = 1200$ ns and $\tau_{\rm th} = 72$ ns. For comparison, a fit of the analytic solution to the same data is shown in \ref{fgr:sim}b. The analytic solution yields $\tau_{\rm gas} = 1260$ ns and $\tau_{\rm th} = 71$ ns. The difference between numerical simulation and analytic fit is 5\%, and the numerical simulation includes the primary resonance peak.

\section{Thermal time constant}
\begin{figure}[ht]
  \includegraphics[width=0.4\linewidth]{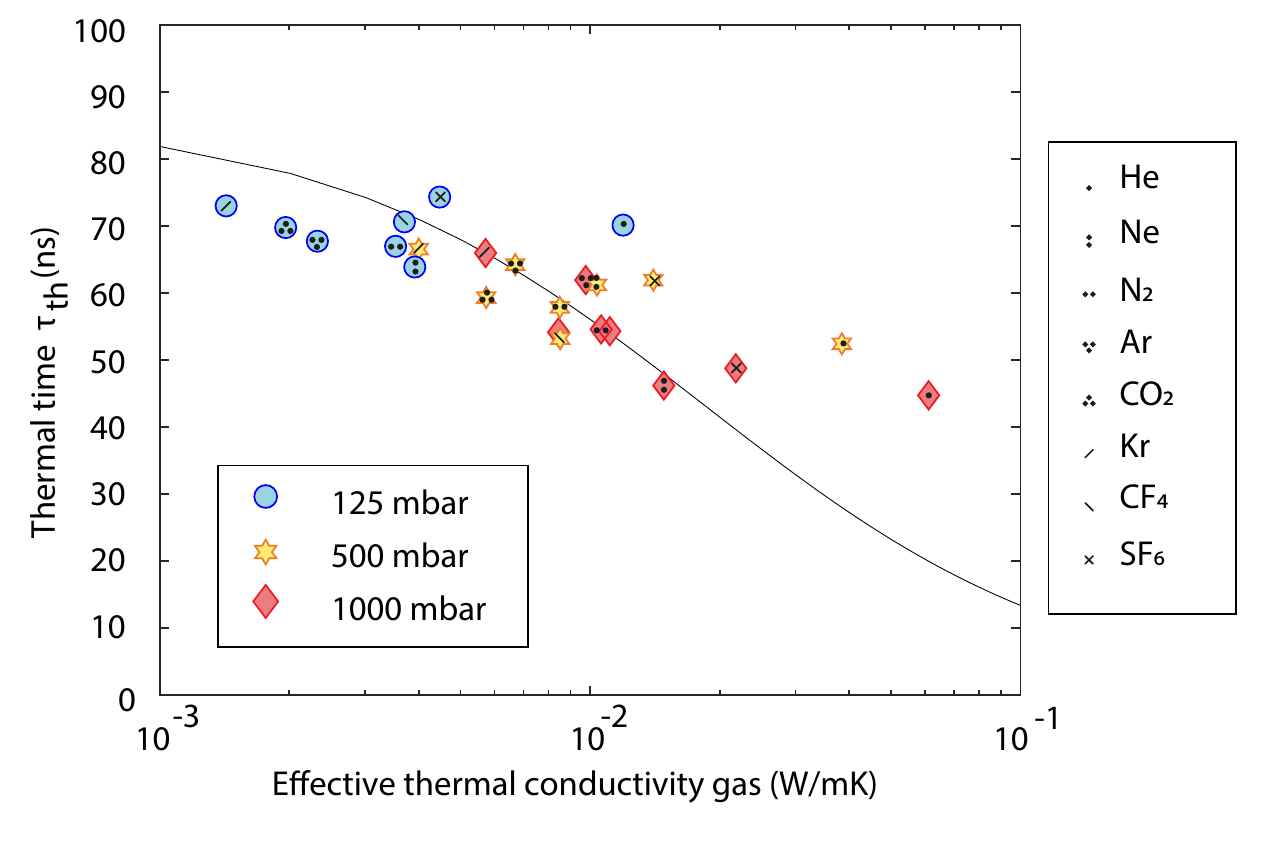}
  \centering
  \caption{The gas offers a new pathway for heat to escape from the membrane, in consequence lowering the thermal time $\tau_{\rm th}$ as the effective thermal conductivity of the gas increases. The black line is a fit to equation \ref{eq:tauth_combo} with fit parameter $\xi = 0.17$. }
  \label{fgr:thermal}
\end{figure}

It is interesting to also investigate the thermal time constant for the different gases at varying pressures. The presence of gas in the cavity opens a new thermal conduction pathway for the membrane and the thermal time constant is therefore expected to decrease as compared to the vacuum measurement. In view of the small dimensions of the gap between the membrane and the substrate the Knudsen formula is used to calculate the effective thermal conductivity $k_{\rm eff}$ of the gas: 
\begin{equation}
\frac{k_{\rm eff}}{k_0} = \frac{1}{1+ 2 \beta {\rm Kn}}.
\end{equation}
Here, $k_0$ is the thermal conductivity of the gas and $\beta$ a constant with a value of about 1.5 that depends on the accommodation coefficient.\cite{reichenauer2007relationship,FluidFlow} Both conduction to the substrate and through the gas contribute to the final thermal time-constant:
\begin{equation}
    \tau_{\rm th}^{-1} = \tau_{\rm th, vac}^{-1} + \frac{\xi k_{\rm eff}}{\rho c_p h_g d}.
    \label{eq:tauth_combo}
\end{equation}
Here, $h_g$, $\rho$ and $c_p$ are the radius, height, density and thermal capacity of the graphene membrane, and $d$ is the cavity depth. A measurement in vacuum is performed to find the thermal equilibration time $\tau_{\rm th, vac} = 87$ ns, which is comparable to values reported in literature for single layer graphene, \cite{dolleman2017optomechanics} suggesting that similar boundary effects are limiting thermal conduction. The constant $\xi$ is a transmission coefficient arising from temperature slip on the solid-gas interface \cite{singh2009modeling}. Figure \ref{fgr:thermal} shows that the gas indeed provides a new heat conduction pathway, decreasing the thermal time constant as effective thermal conductivity increases. From the data a value of $\xi = 0.17$ is found to fit our experiments.

Gas sensing can be achieved by observing changes in the thermal time constant in a fashion similar to Pirani gas sensors. In general the gas conducts heat better at higher pressures, and it does so also for molecules with a smaller molecular mass and higher molecular velocity. However, it appears that thermal conductivity of the gases is a less precise route toward gas sensing than the permeation based method discussed in the main text.

\newpage
\section{Single layer graphene circular drum with nanoperforations}
Smaller perforations with sizes below one nanometer could enable molecular sieving and enhance responsivity of these devices. With this purpose, some single layer graphene drums have been exposed to highly energetic ion bombardment with 129Xe$^{23+}$ 0.71 MeV/u, with a flux ranging from $5.09 \cdot 10^7$ to $5.09 \cdot 10^9$ ions per square centimeter at the SME beamline of GANIL (Caen, France). This is similar to the treatment described by Madauß et al.\cite{madauss2017fabrication} Characterization of the nano indentations on the drum is performed using AFM. This experiment is of interest since it shows that our gas sensing principle works using a single layer circular membrane with defects which could potentially lead to applications benefiting from molecular sieving.

\begin{figure}[H]
  \includegraphics[width=0.6\linewidth]{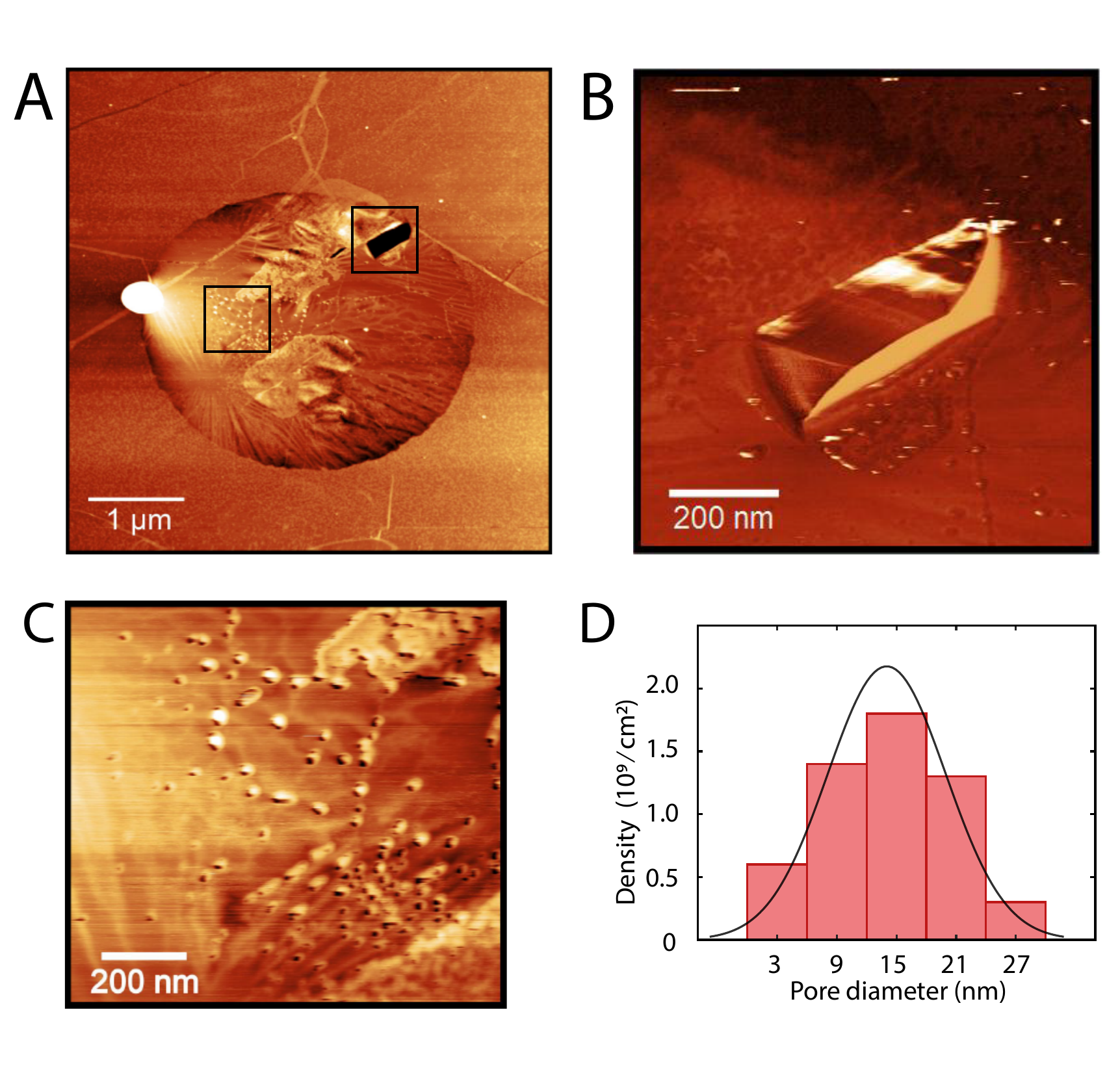}
  \centering
  \caption{a) Image of the drum using AFM. The white spot on the left is a particle of presumably dust. b) The rip in the membrane is pictured using the phase channel of the AFM. The ripped graphene sheet is still attached to the membrane. c) Detail with the nano pores visible. d) The size and number of pores has been counted in this area. The sizes are distributed normally with mean 14 nm.}
  \label{fgr:AFM}
\end{figure}

\begin{figure}[H]
  \includegraphics[width=0.8\linewidth]{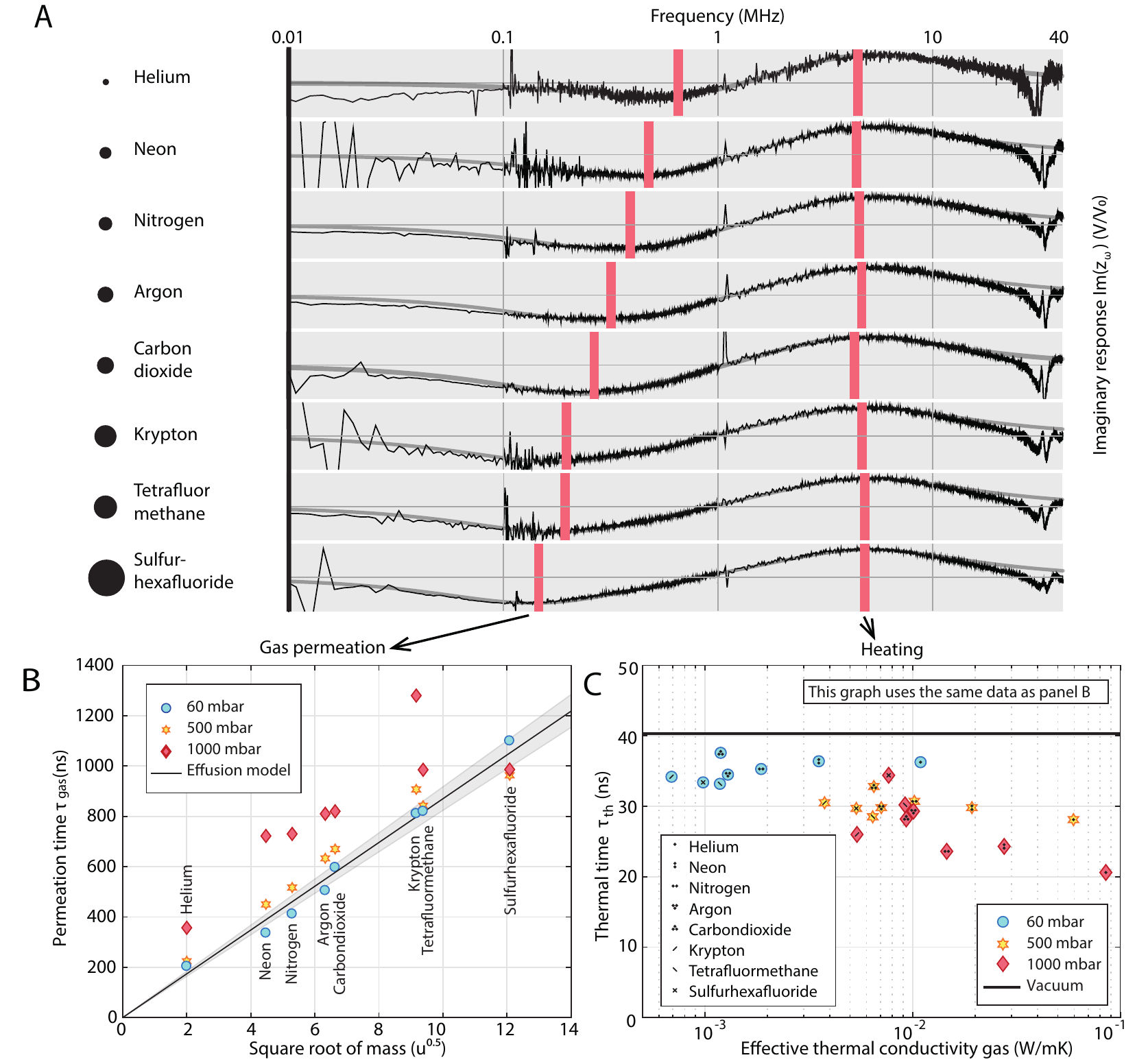}
  \centering
  \caption{a) Measurements on a circular SLG drum, which is depicted in figure \ref{fgr:AFM}, show the same characteristics as the dumbbell shaped devices. b) Permeation time constants follow Graham's law. c) Thermal time constants become lower with higher effective thermal conductivity.}
  \label{fgr:AFMmeas}
\end{figure}

\bibliographystyle{apsrev4-2}
\bibliography{references}

\end{document}